\documentclass[twocolumn,preprintnumbers,amsmath,amssymb]{revtex4}
\usepackage{graphicx}
\usepackage{color,soul}
\usepackage{amssymb,amsfonts,amsmath,wasysym,cancel}
\usepackage{setspace}
\usepackage{lpic}

\newcommand{\nl}{\nonumber \\}

\newcommand{\vf}{v_{\rm f}}
\newcommand{\vft}{\tilde{v}_{\rm f}}
\newcommand{\zc}{z_{\rm c}}

\newcommand{\gammac}{\gamma_{\rm c}}
\newcommand{\E}{G} 
\newcommand{\Eo}{G_0} 
\newcommand{\dz}{{\Delta z}} 
\newcommand{\ts}{{t_c}} 
\newcommand{\V}{{\hat v}} 
\newcommand{\X}{{\hat x}} 
\newcommand{\T}{{\hat t}} 
\newcommand{\Vo}{{\V_0}} 
\renewcommand{\S}{{\hat s}} 

\DeclareMathOperator{\erfc}{erfc}

\begin{document}

\title{Shear shocks in fragile networks}

\author{S. Ulrich, N. Upadhyaya, B. van Opheusden and V. Vitelli}
\affiliation{Instituut-Lorentz for Theoretical Physics, Leiden University, NL 2333 CA Leiden, The Netherlands}



\begin{abstract}

A minimal model for studying the mechanical properties of amorphous solids is a disordered network of point masses connected by unbreakable springs.  At a critical value of its mean connectivity, such a network becomes fragile: it undergoes a rigidity transition signaled by a vanishing shear modulus and transverse sound speed. We investigate analytically and numerically the linear and non-linear visco-elastic response of these fragile solids by probing how shear fronts propagate through them. Our approach, that we tentatively label shear front rheology, provides an alternative route to standard oscillatory rheology. In the linear regime, we observe at late times a {\it diffusive} broadening of the fronts controlled by an effective shear viscosity that diverges at the critical point. No matter how small the microscopic coefficient of dissipation, strongly disordered networks behave as if they were {over-damped} because energy is irreversibly leaked into diverging non-affine fluctuations. Close to the transition, the regime of linear response becomes vanishingly small: the tiniest shear strains generate strongly non-linear shear shock waves qualitatively different from their compressional counterparts in granular media. The inherent non-linearities trigger an energy cascade from low to high frequency components that keep the network away from attaining the quasi-static limit. This mechanism, reminiscent of acoustic turbulence, causes a super-diffusive broadening of the shock width.

\end{abstract}

\maketitle

\section{Introduction}

Many natural and man-made amorphous structures ranging from glasses to gels can be modeled as disordered viscoelastic networks of point masses (nodes) connected by springs. Despite its simplicity, the spring network model uncovers the remarkable property that the rigidity of an amorphous structure depends crucially on its mean coordination number $z$, i.e., the average number of nodes that each node is connected to \cite{Alexander}. For an unstressed spring network in $D$ dimensions, the critical coordination number $\zc=2D$ separates two disordered states of matter: above $\zc$ the system is rigid, below $\zc$ it is floppy. Therefore, $\zc$ can be identified as a critical point in the theory of rigidity phase transitions \cite{PhysRevB.31.276,Ohern,sun2012}. 

Various elastic properties are seen to scale with the control parameter $\dz=z-\zc>0$, close to the critical point \cite{Alexander}. For example, the shear modulus vanishes as a power law of $\dz$ \cite{ellenbroek2009,zaccone2011,Thorpe}. At the critical point, the disordered network becomes mechanically fragile in the sense that shear deformations cost no energy within linear elasticity . This property is shared with packings and even chains of Hertzian grains just in contact, which display also a zero bulk modulus \cite{Durian,Ohern}. There is however an important difference. In the disordered spring networks local particle interactions are harmonic. Fragility and the incipient non-linear behavior is a collective phenomenon triggered by the mean global topology of the network which is composed of unbreakable Hookean springs \footnote{No smooth deformation can change $z$ -- springs must be added or removed.}. By contrast, in granular media, additional and crucial non-linearities are typically introduced also at the local level, eg. by the Hertzian interaction between grains 
or particle rearrangements that leads to a distinct shock phenomenology \cite{Gomez2012,PhysRevE.86.041302,Nesterenko,waitukaitis2013dynamic,waitukaitis2012impact,Martin}.    

In the present work, we adopt a blend of theoretical and numerical analysis to study the out of equilibrium response of viscoelastic random spring networks subjected to a constant influx of energy. Our approach suggests an alternative route to standard oscillatory rheology that consists in constantly shearing one of the boundaries, while monitoring how the ensuing shear front propagates throughout the sample, see Fig. 1. This method, that we tentatively label shear front rheology (SFR), is of general applicability, but it may be particularly appropriate to investigate critical systems whose dynamics is undergoing a dramatic slow-down, such as biopolymer networks or gels with a sufficiently small shear modulus and sound speed, so that transients become the main observables. Our findings demonstrate the simultaneous break-down of three widely adopted, but dangerous, assumptions: at the the critical point, the mechanical response is inherently non-affine, non quasi-static and non-linear. 

\begin{figure}
 \centering
  \includegraphics[width=.49\textwidth]{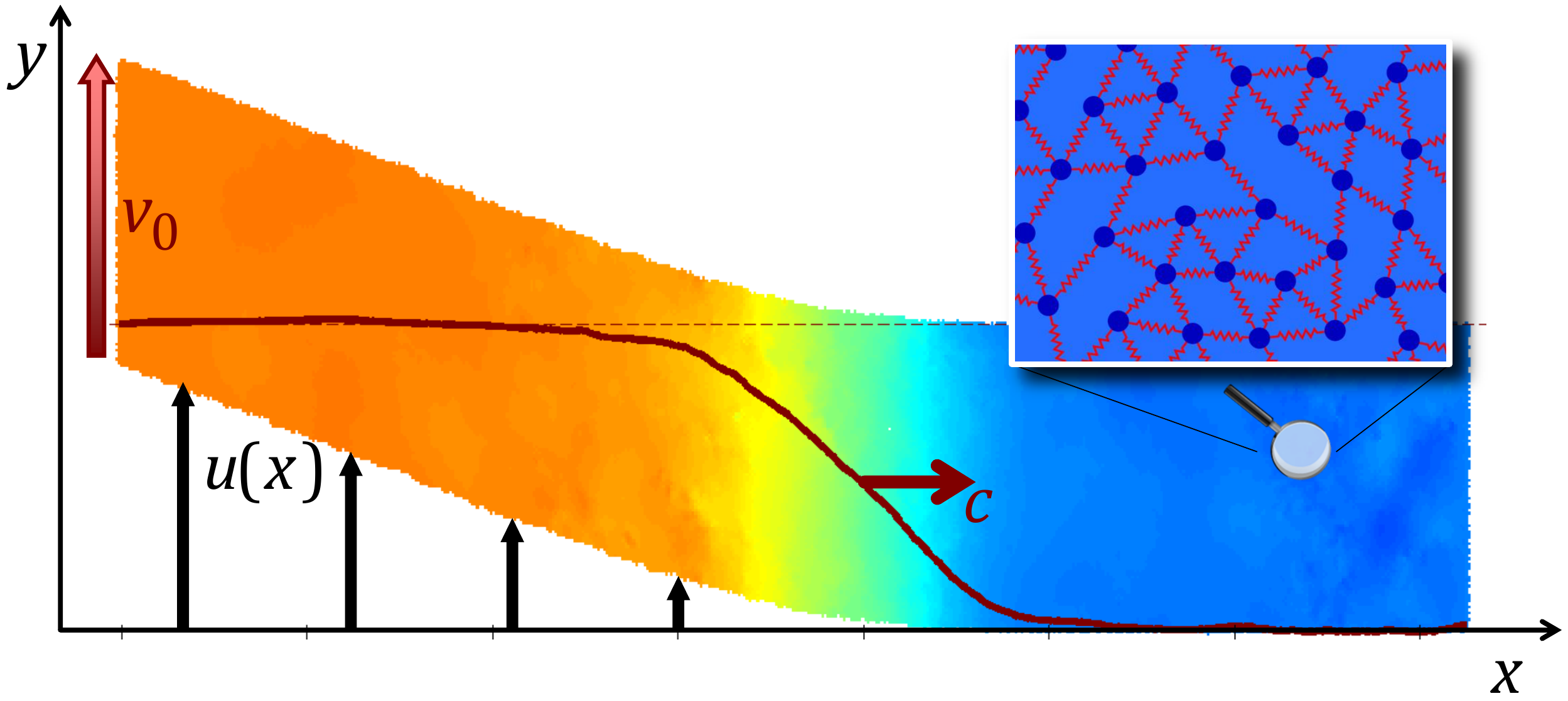} 
 \caption{Simulation snapshot of a shear front moving with speed $c$ in a network under shear. The left boundary is sheared upwards with a constant velocity $v_0$, while the right boundary is kept fixed. The color encodes the velocity in the $y$-direction at each point in the sample (orange means moving upwards and blue means stationary). The displacement field $u(x)$,  depicted as black arrows, has a nearly constant downwards slope behind the front, while it is approximately zero ahead of it. Superimposed as dark red line is the $y$-component of the actual velocity field $v(x)$, after taking the average in $y$-direction (transverse direction). It is normalized such that the dashed line is $v_0$. The inset shows a magnified portion of the disordered network.}
 \label{fig:systemSchematic}
\end{figure}

First, the approach towards the critical point is accompanied microscopically by an increasingly heterogeneous displacement field that is non affine with respect to the uniform shear applied at the boundary \cite{zaccone2011,Zacconeb,Ganguly,PhysRevLett.93.195501,Wyart2008,PhysRevE.80.061307}. We show that the increasingly non-affine response, as $\Delta z$ is lowered towards the rigidity transition, is reflected in SFR by the diverging widths of the shear fronts. This effect is a dynamical analogue of the diverging width of magnetic domain walls near the Curie point -- it is captured in our elastic models by a divergent disorder-induced effective viscosity. Thus, even in the limit of vanishing microscopic coefficient of viscous dissipation, a disordered spring network behaves as if it were over-damped. A large fraction of the energy injected at the boundary is leaked into non-affine fluctuations and unless a constant influx of energy is maintained at the boundaries, coherent wave propagation cannot be observed. This effect is reminiscent of the strong ultrasound attenuation in smectic liquid crystals, where the role of the soft non-affine fluctuations triggered by disorder is played by soft layer-bending fluctuations, at finite temperature \cite{Toner}. 

Second, we find that, no matter how slowly you shear, the process can never be considered quasi-static at the critical point. In oscillatory rheology, this means that the frequency range $\omega < \omega_c (\dz)$ below which the network must be sheared to observe a frequency independent (dc) response, becomes vanishingly small (eg. $\omega_c \rightarrow 0$), as the critical point is approached \cite{tighe2011}. The progressive breakdown of the quasi-static approximation is signaled in SFR by the front width broadening super-diffusively with time. We show, that from the the super-diffusion exponent in SFR you can read off the power law exponent of the loss modulus, as a function of frequency, in oscillatory rheology. In this regime, the network has no time to respond elastically so there is no front propagation only the super-diffusive broadening that would be also observed for $\Delta z < 0$. Away from the critical point, simple diffusive broadening of the front is recovered after a divergent time scale that is inversely proportional to $\omega_c (\dz)$.

Third, close to the critical point the shear modulus and transverse sound speed become vanishingly small, as the regime of linear response progressively vanishes \cite{Xu}. Thus, at the rigidity transition, shear deformations can no longer propagate as linear shear waves because the structure is mechanically fragile. Instead, we demonstrate that even for the tiniest strains, non-uniform strongly non-linear shock waves arise. Away from the critical point, linear shear waves are restored but only below a critical strain, $\gamma_c$, that vanishes at the transition. For strains larger than $\gamma_c$, the same shear shocks that emanate from the critical point are responsible for the transmission of mechanical energy. We show that the speed at which these non-linear fronts move can serve as a dynamical probe of the non-linear exponent in the stress-strain relation. Unlike linear fronts, the shock width broadens super-diffusively (even at late times) as a result of a non-linear energy cascade reminiscent of acoustic turbulence. This mechanism acts as a source that generates higher and higher harmonics that keep the network away from attaining the quasi-static limit. 

Experimental and theoretical investigations of compressional shocks and solitons have been carried out extensively in granular media prepared in a state of zero external pressure (termed sonic vacuum) in which the bulk modulus vanishes as a result of non-linear grain interactions \cite{Nesterenko, Gomez2012,Martin}. By contrast, shear shocks triggered by global non-linearities are less explored experimentally with the notable exception of bio gels that share with our models the property of having a very small shear modulus typically of order of kPa \cite{PhysRevLett.91.164301}, but they are typically stabilized by osmotic pressure or activity \cite{Alexander,Fred}. Nonetheless, it is much easier to break the sound barrier (and observe genuine {\it acoustic} Mach cones) in human tissues, where the speed of linear transverse sound is of the order of $m/s$, than in metals where it is thousands times higher. Indeed, supersonic shear imaging has recently developed into a powerful diagnostic tool for medical applications \cite{1320804,fink:28}.   


\section{Divergent shear front width}

We construct computer models of weakly connected two-dimensional random viscoelastic networks from highly compressed jammed packings of frictionless poly-disperse disks \cite{Ohern,somfai2005}. One first identifies the disk centers as point particles (network nodes) and then models the  interactions between overlapping disks using two sided harmonic springs of varying rest length to eliminate any pre-stress existing in the original jammed packings. The result is a highly coordinated ($z \approx 6$) spring network that serves as the seed from which families of networks with a wide range of $z$ are generated by removing springs. We have investigated two cutting protocols that result in two distinct ensembles of random spring networks. 

The first protocol generates nearly-homogeneous isotropic networks by progressively removing bonds with the highest $z$, such that spatial $z$ fluctuations are reduced by construction, a scenario assumed in several mean field theories of the jamming transition \cite{Zamponi_RMP,Wyart2008}. The second protocol consists in removing springs \emph{randomly} (subject to the no-local-floppiness constraint $z>3$) and results in less homogeneous but still isotropic structures, whose critical exponents are close to the ones obtained in rigidity percolation \cite{Thorpe}. These two ensembles of networks display different critical exponents, but share the crucial qualitative features that arise from the presence of a rigidity transition at $z_c$. For simplicity, we present most of our results in the context of the homogeneous networks while noting how the exponent are modified if random cutting is used.  

 \begin{figure}[h]
\centering
\begin{picture}(290,188)
    \put(0,0){\includegraphics[width=.49\textwidth]{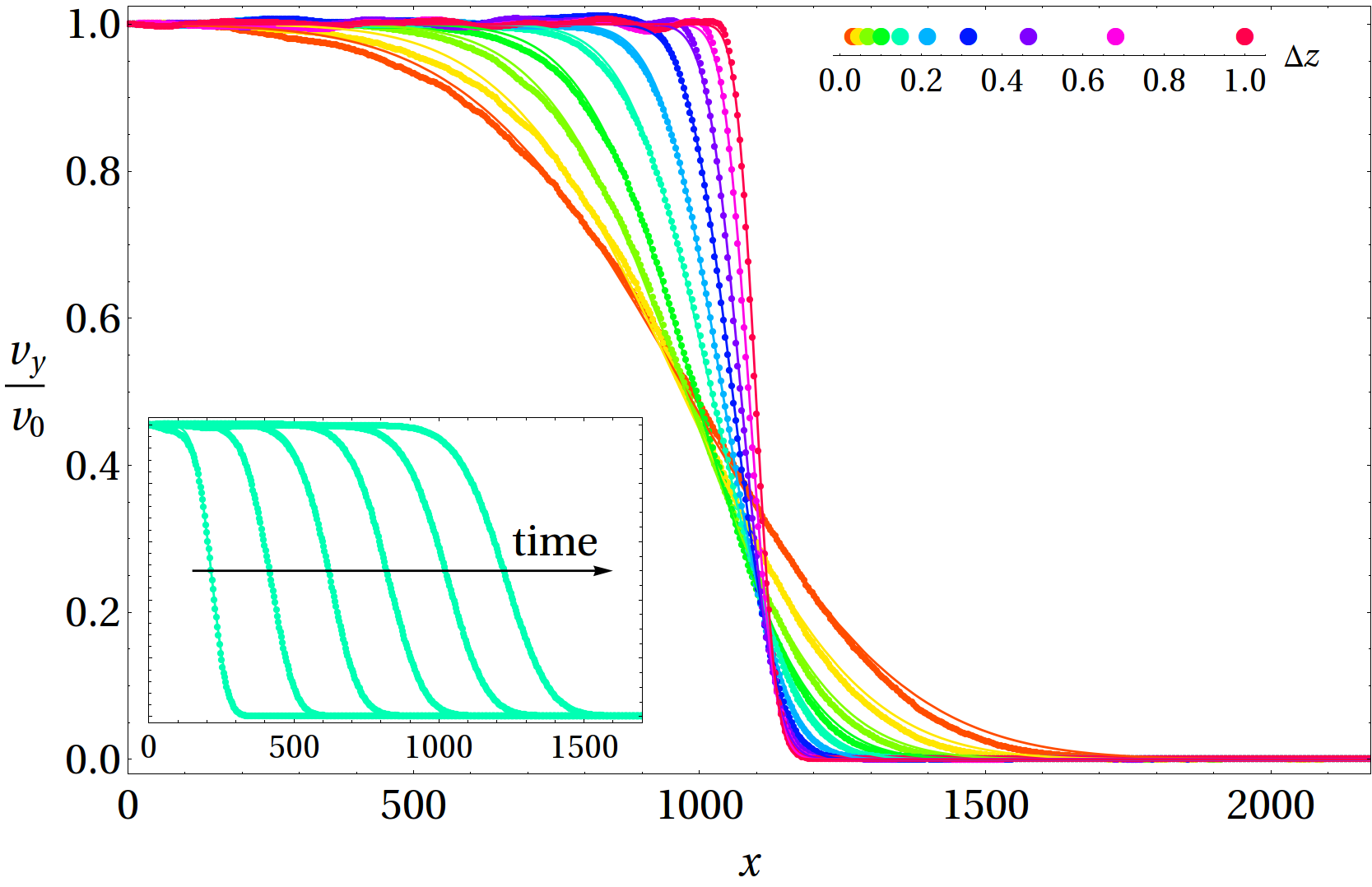}}
    \put(145,70){\includegraphics[width=.2\textwidth]{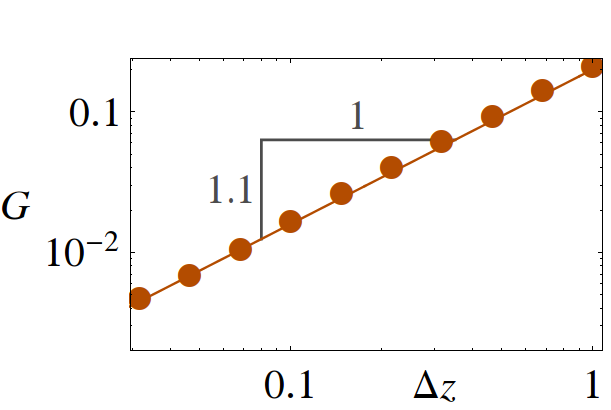}}
  \end{picture}
\caption{Shape of the linear wave profile. In the main plot, velocity profiles for different $\dz$ are superimposed, with the smallest $\dz$ corresponding to the widest profile. The damping coefficient is $b=0.1$ and the time is chosen such that the wave has roughly reached the center of the sample. The left inset shows the wave front for $\dz = 0.15$ (same color as main plot) for different times. 
In all plots, data points are (averaged) profiles from the simulation and the solid lines are fits to Eq.~\eqref{eq:apndx:V(XT)1F1}. The right inset shows the shear modulus $G$ versus $\dz$ extracted from the speed of the linear front propagating in a homogeneously cut network.}. 
 \label{fig:timeDependence}
\end{figure}

Once the networks are generated, we shear the left most edge at a constant speed $v_0$, as in Fig.\ 1, and follow the evolution of the resulting shear velocity profile by averaging out the longitudinal particle speeds over bins along the $x$ direction effectively creating a one dimensional front profile propagating in the $x$ direction. The dynamics is obtained by numerically integrating Newton's equations of motion (using the velocity Verlet method) subject to Lees-Edwards boundary conditions in the $y$-direction and hard walls in the $x$-direction \footnote{A hard wall can only move up and down as a whole, but no relative motion of the particles is allowed.}. The samples are composed of up to $N=10^5$ identical particles  with mass $m$. In addition to the Hookean interaction (with a spring constant $k$) between connected particles, $i,j$,  we include the effects of viscous dissipation: $\vec{f}_\text{diss}^{ij} = - b  (\vec{v}^{i} - \vec{v}^{j})$, where $b$ is the microscopic damping constant, and $\vec{v}^{i,j}$ are the velocities of a pair of particles connected by a spring. Time is measured in units of $\sqrt{m/k}$ and the damping constant in units of $\sqrt{km}$, which is equivalent to setting $m=1$ and $k=1$. Lengths are measured in units of $d$ the mean spring length at rest. 
 
In Fig.\ 2,  we illustrate the late-times dynamics of the transverse velocity profiles $v(x,t)$, normalized by the shearing speed $v_{0}$, for a small shear strain $\gamma = 10^{-3}\gammac$, where $\gamma_c$ is the critical strain needed to elicit a non-linear response, discussed in detail later. Inspection of the schematics in Fig.\ 1 shows how the shear strain, $\gamma \equiv {\partial u}/{\partial x}$, is implicitly controlled by $v_{0}$ via the relation $\gamma={v_0}/{v_\text{f}}$. The bottom left panel shows successive snapshots of the (transverse) velocity profiles $v(x,t)$ as a function of $x$ (after averaging the longitudinal velocity fluctuations) in a sample at $\Delta z=0.15$. After an initial transient, a linear shear front propagates at a constant transverse sound speed, $c = \sqrt{{G}/{\rho}}$, where $\rho$ is the mass density and $G$ the shear modulus. 
In the right inset of Fig.\ 2, the numerical value of $G$ extracted from the speeds of shear fronts propagating in homogeneously cut networks is plotted against $\dz$ (red dots). We find that it is consistent with the expected linear scaling $G \propto \dz$ \cite{Ohern,Durian,ellenbroek2009} (the best fit solid red line has a slope close to 1.1). By contrast, a power law exponent numerically closer to $1.4$ is found for randomly cut networks.

In the main panel of Fig.~2, we show the velocity profiles obtained from molecular dynamics simulations after the fronts have travelled half the length of homogeneous random networks with different $\dz$ (the numerical data are represented by symbols with different colors). In order to unambiguously extract the dependence of the width $w(t)$ on $\dz$, we first fit the data with an analytical solution for the velocity profile derived in the Supplementary Information. Next, we define the rescaled width $\tilde{w}^2(\dz)\equiv w^2(t)/t$ that normalizes out the diffusive spreading in time and plot $\tilde{w}$ as a function of $\dz$,  as red dots in the main panel of Fig.\ 3. The solid line that fits the data for $\tilde{w}$ has slope $-1/2$ on a log plot, hence $\tilde{w} \propto \dz ^{-1/2}$ for homogeneously cut networks. The divergent width of the shear fronts is a striking signature (in real space) of the important diverging length scale proportional to  $\dz ^{-1/2}$ that accompanies the jamming/unjamming transition \cite{PhysRevLett.95.098301,PhysRevE.81.021301,ikeda,during2012}. 

Note, however, that the scaling exponent of the width $\tilde{w} \propto  \dz^{-\alpha}$ (as for the shear modulus one) is not universal, instead it depends on the spring cutting protocol adopted to generate the network. If springs are cut randomly, we find that $\alpha$ is numerically close to $3/4$, see red data in the lower inset of Fig.\ 3, but the basic phenomenology is essentially unaltered. The divergence of the front width as a function of $\dz$ occurs irrespective of the choice of the microscopic coefficient of dissipation even when $b=0$, provided that disorder is strong (see green data in the lower inset of Fig.\ 3). As we shall see in the next section, the divergent front width results from the divergence of non-affine fluctuations at the critical point -- it can be modeled analytically upon introducing a divergent effective viscosity. 

\begin{figure}
\noindent
\includegraphics[width=.49\textwidth]{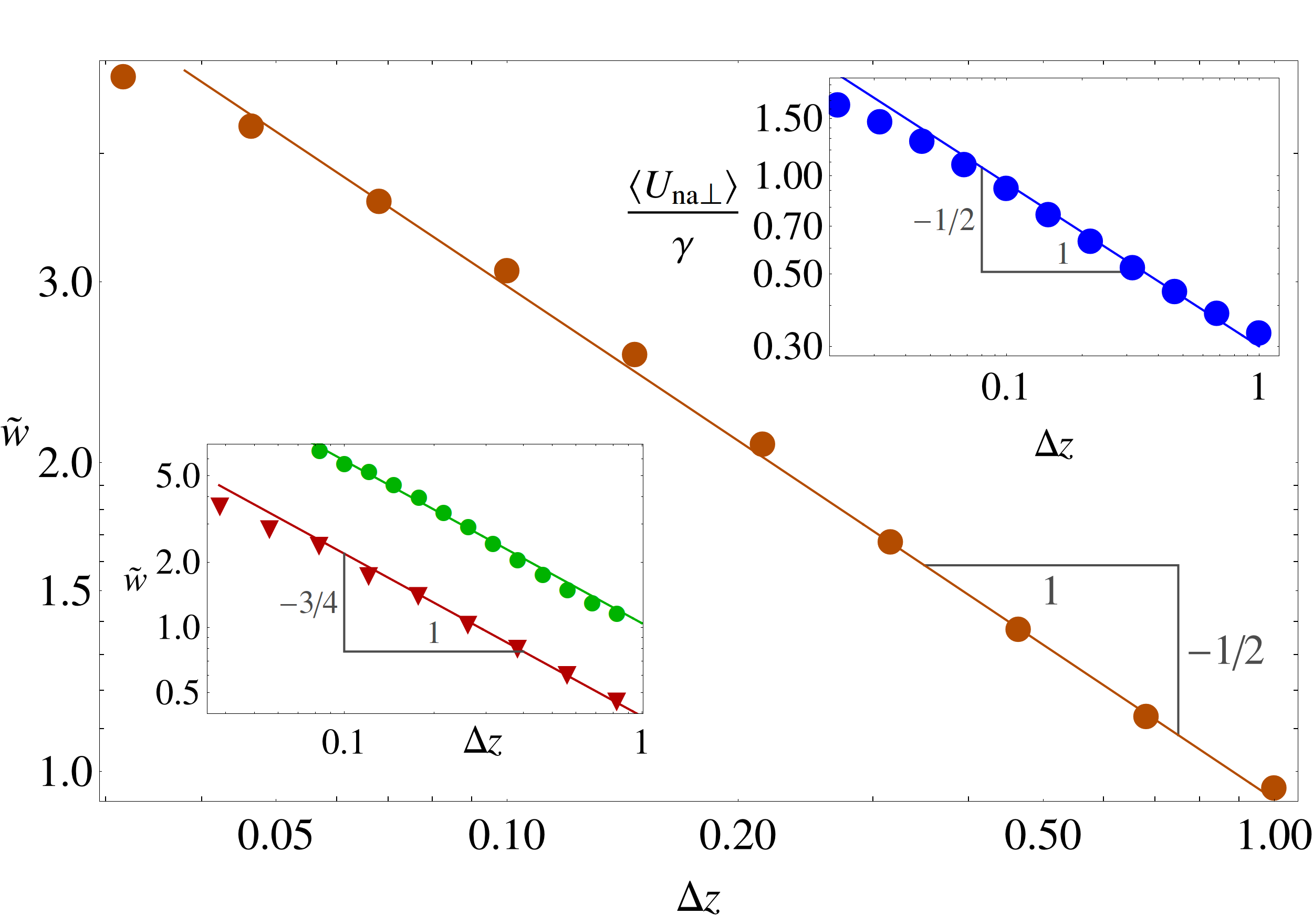}
 \centering
\caption{The main panel shows numerical data for the rescaled width $\tilde{w} \equiv w(t)/\sqrt{t}$ (or equivalently the effective viscosity $\eta_e$) versus $\dz$ (dots) for homogeneously cut networks with $b=01$. The solid line represents the scaling $\tilde{w} \propto \dz^{-1/2}$. 
The right inset shows the average of the non-affine displacements, $U_{na}$, where only the components perpendicular to the respective bond vector have been taken. The data has been normalized with the shear strain $\gamma$. The solid line has the form $U_{na} \propto \dz^{-1/2}$ and underlines the divergence of the non-affine deformations for $\dz \to 0$. The left inset shows $\tilde{w} \equiv w(t)/\sqrt{t}$ versus $\dz$ for randomly cut networks with $b=0.1$ (blue dots) and $b=0$ (red dots). The continuous lines have slope $-3/4$ which closely approximates the data.
} 
\label{fig:ShearModuliAndViscosities}
\end{figure}

\section{Shear front rheology and effective viscosity} 

Consider first, the general  linear visco-elastic relation between the stress $\sigma \equiv \sigma_{xy}$, the strain $\gamma \equiv \gamma_{xy}$ and the strain rate $\dot{\gamma} \equiv \frac{\partial \gamma}{\partial t}$ in the frequency independent (quasi-static) regime
\begin{equation}
\sigma = G \gamma + \eta_e \dot{\gamma}  \;, \label{eq:zero}
\end{equation}
where the loss modulus $\eta_e$ (viscosity) and shear modulus $G$ are constants, independent of frequency. For random networks however, these moduli can be scaling functions of $\dz$ that we denote by {$G \propto \dz^{\beta}$} and $\eta_{e} \propto \dz^{-\chi}$, respectively. Upon expressing the shear strain as $\gamma = \frac{\partial u}{\partial x}$ (see Fig.\ 1), we write down Newton's equation as
\begin{equation}
\frac{\partial \sigma}{\partial x} = \rho \frac{\partial^2 u}{\partial t^2} = G \frac{\partial^2 u}{\partial x^2} + \eta_e \frac{\partial^3 u}{\partial x^2 \partial t}  \;, \label{eq:motion}
\end{equation}
This linear second order partial differential equation can be readily cast in terms of the velocity field $v(x,t)\equiv \frac{\partial u(x)}{\partial t}$. Inspection of Eq.~\eqref{eq:motion} reveals that the resulting velocity profile propagates as a linear wave with speed proportional to $\sqrt{G/\rho}$ (from the second order terms), while its width spreads diffusively as $w^2 \propto \eta_e t$ (from the third order term). Hence, the $\tilde{w} \propto \dz ^{-\alpha}$ scaling observed in the fluctuations-averaged $v_{y}(x)$ profiles of Fig.\ 2, implies that $\eta_e \propto \dz^{-\chi}$ with $\chi=2\alpha$. 

In order to uncover the physical origin of the diverging viscosity $\eta_e$, note that the displacement field $u(x)$ in Eq. \ref{eq:motion} describes only the dynamics of the affine deformations that the network undergoes. However, the relative magnitude of the non-affine fluctuations $\langle U_{na} \rangle$, that have been averaged out in Fig.\ 2, diverges while approaching the critical point as $\langle U_{na}\rangle \propto \dz^{-1/2}$ for homegeneously cut network \cite{ellenbroek2009,Wyart2008}, as verified in Fig.\ 3 top inset. In our treatment, we implicitly capture the presence of the large non-affine fluctuations by a renormalization of the effective viscosity $\eta_e \propto \dz ^{-\chi}$ that diverges as the critical point is approached (even if $b=0$). The scaling exponent $\chi$ (that is equal to $1$ for homogenous network) is suitably chosen to match the amount of kinetic energy $\dot{U}^2_{na}\propto \dz^{-1} $ leaked into the non-affine fluctuations that have been integrated out.  

\begin{figure}
\noindent
 \includegraphics[width=.49\textwidth]{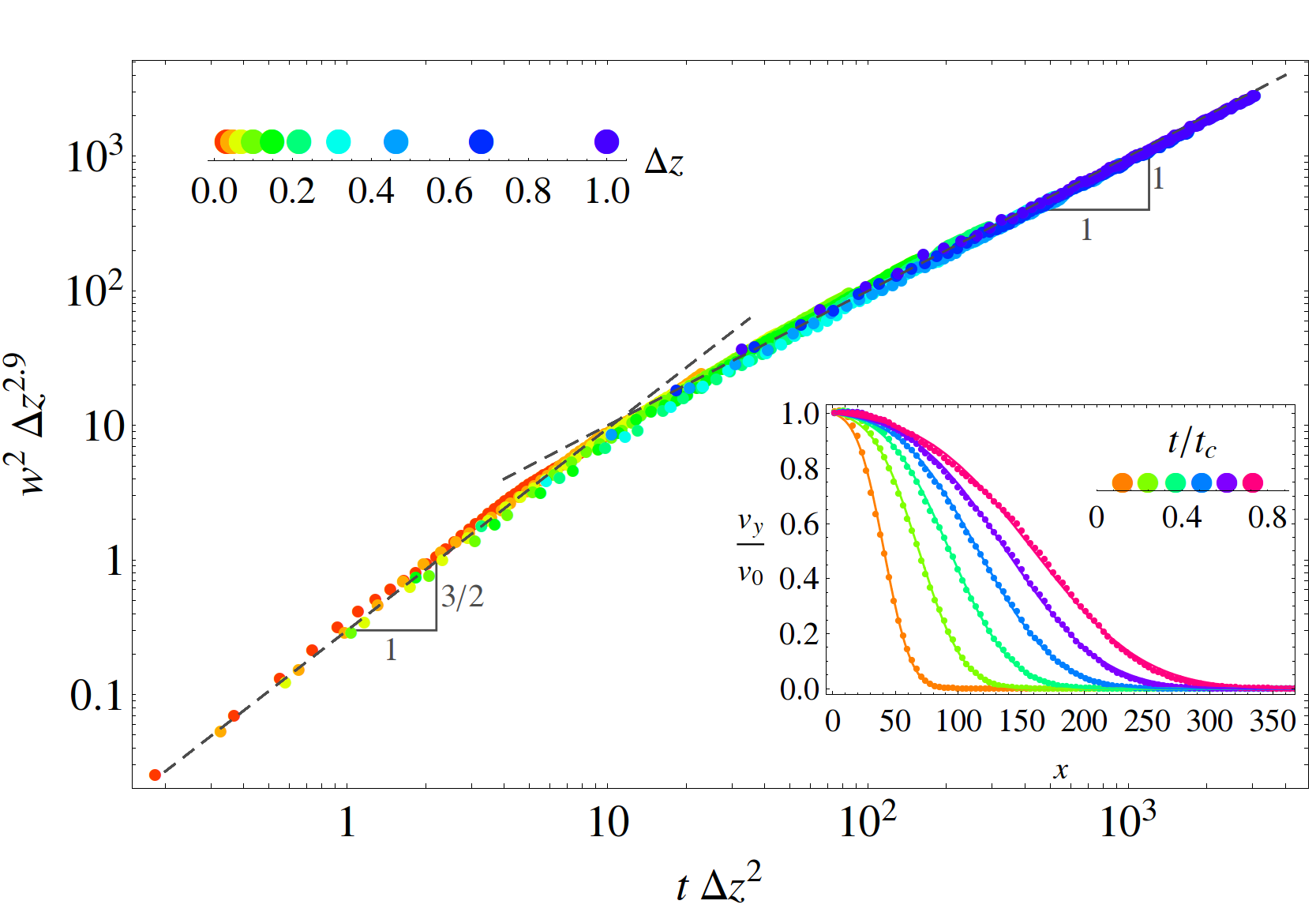}
 \centering
\caption{The main plot shows the time evolution of the squared front width, whereby the time axis has been normalized by $\ts \propto \dz^{-2}$ and the width axis by $\dz^{2.9}$. The inset shows the broadening at early times, $t < t_c$, of the velocity profiles, $v(x,t)$, in the $\hat{y}$ direction, normalized by $v_0$. Note the absence of front {\it propagation}, in contrast with the plot in the left inset of Fig.\ 2.}
\label{fig:widthVsTime}
\end{figure}

By assuming that $G$ and $\eta_e$ are frequency independent, we restrict the domain of validity of our predictions to the so called \emph{quasistatic regime}. The shear fronts enter this regime only after a characteristic time scale $t_c$ that diverges at the transition. The $\dz$ dependence of $t_c$, derived in the Supplementary Information, can be guessed from the ratio ${\eta_e}/{G}$ which for homogeneously cut networks gives $t_c \propto \dz^{-2}$, while the randomly cut network are found numerically to have a scaling exponent close to $-3$. For $t \ll t_c$, the network has ``no time'' to respond elastically so we are effectively probing the properties of a floppy structure, even if $z>z_c$. The experimental signature of this regime is that the shear deformation continuously applied at the edge penetrates the sample mainly by broadening its width instead of propagating, as illustrated in the inset of Fig.\ 4. 

In oscillatory rheology, the $t \ll t_c$ regime corresponds to the non-quasistatic regime $\omega \gg \omega_c$ (with $\omega_c \propto {t_c}^{-1}$), for which the viscosity is no longer frequency independent, but can exhibit the power law scaling $\eta_e \propto \omega^{-\delta}$ \cite{tighe2011}. Heuristic considerations (formalized in the Supplementary Information) suggest that the front width $w^2 \propto \eta_e t \propto t^{1+\delta}$ broadens super-diffusively for $t \ll t_c$. The super-diffusive spreading is corroborated in the main panel of Fig.\ 4, where the width $w(t)$ is shown as a function of $t/t_c$ for a range of $\dz$, indicated by different colors. From the measured super-diffusion exponent $w^2 \propto t^{3/2}$ we infer that $\eta_e(\omega) \propto \omega^{-1/2}$ (independently of cutting protocol): this scaling is often encountered in rheological studies of grains and emulsions \cite{PhysRevLett.76.3017,tighe2011}. For $t \gg t_c$ the fronts propagate ballistically and their widths broaden diffusively with an $\omega$ independent effective viscosity $\eta_e \propto \dz^{-1}$, for homogeneously cut networks. Consistently, all data can be collapsed onto a single master curve when $w$ is rescaled by a power of $\dz$
close to 3 (note $w^2 \dz^3 \propto t \dz^2$ implies $\eta_e \propto \dz^{-1}$). 

The more general response function treatment, detailed in the Supplementary Information, formalizes how to extract the full $\omega$ dependence of the moduli from the velocity profiles. Consider the frequency dependent constitutive relation for a linear visco-elastic material
\begin{equation}
\sigma(s) = G(s) \gamma(s) \;, \label{eq:sigmaGeneralLaplace}
\end{equation}
where  $G(s)_{s=i\omega}=G'(\omega) + iG''(\omega)$ is the (Laplace transformed) complex modulus, whose real and imaginary parts correspond (aside from an $\omega$ factor) to the shear (storage) and viscosity (loss modulus) respectively. Assuming that the random spring network is homogeneous and isotropic, we solve the equation of motion  for the velocity field to obtain (see Supplementary Information) 
\begin{equation}
 \frac{v(x,s)}{v_0} =   \frac{1}{s} \exp\!\left( -\frac{\sqrt{\varrho} \, s x}{\sqrt{\E(s)}}  \right) \;, \label{eq:app:v(x,s)}
\end{equation}
where $v_0$ is the strain speed imposed at the boundary (see Fig.~1) and $x$ is a coarse grained variable along the longitudinal direction. 

We can readily invert the Laplace transform to obtain the velocity profile, $v(x,t)$, as a function of time, once an appropriate choice for the complex modulus is made or vice-versa. The quantitative match between the predictions of Eq. (\ref{eq:app:v(x,s)})  (see the continuos lines in the main panel of Fig.\ 2) and our numerical data validates the concept of an effective viscosity. It is a useful tool to capture within ordinary continuum elastic theory the divergent non-affine fluctuations responsible for its breakdown. 

\section{Shear shocks and energy cascade \label{sec:theory}}

As the critical point is approached, the linear approximation, adopted in the previous sections, progressively breaks down because the shear modulus and transverse sound speed become vanishingly small. This raises the question of how energy is transferred in such a fragile material. A simple guess for the non-linear stress strain relation reads
\begin{equation}
\sigma = G \gamma + k_{nl} {\gamma}^n  \;, \label{eq:zerobi}
\end{equation}
where $k_{nl}$ is a non-linear elastic coefficient that does not vanish at $\dz=0$ and $n$ is a non-linear exponent to be inferred from shear front rheology experiments ($n=3$ is the most common result that insures mirror symmetry). 

In most elastic media, the first term in \eqref{eq:zerobi} dominates as long as $\gamma \ll 1$. Thus, the fundamental mechanical excitations are (weakly interacting) transverse phonons. By contrast, in fragile matter, the usual approach needs to be turned around because $G \propto \dz^{\beta}$ is the small parameter. Hence, the basic excitations are now obtained from the non-linear term -- they are shock-like solutions of the equations of motion derived from Eq.~\eqref{eq:zerobi}. By equating the two terms in Eq.~\eqref{eq:zerobi}, we find the threshold $\gammac \sim \dz^{\frac{\beta}{n-1}}$ above which the non-linear response kicks in, which is indeed much smaller than $\cal{O}$$(1)$ close to the transition.     

\begin{figure*}
\noindent \hspace{-6pt}
\begin{minipage}[h]{.59\textwidth}
  \begin{picture}(290,210)
    \put(0,0){\includegraphics[width=\textwidth]{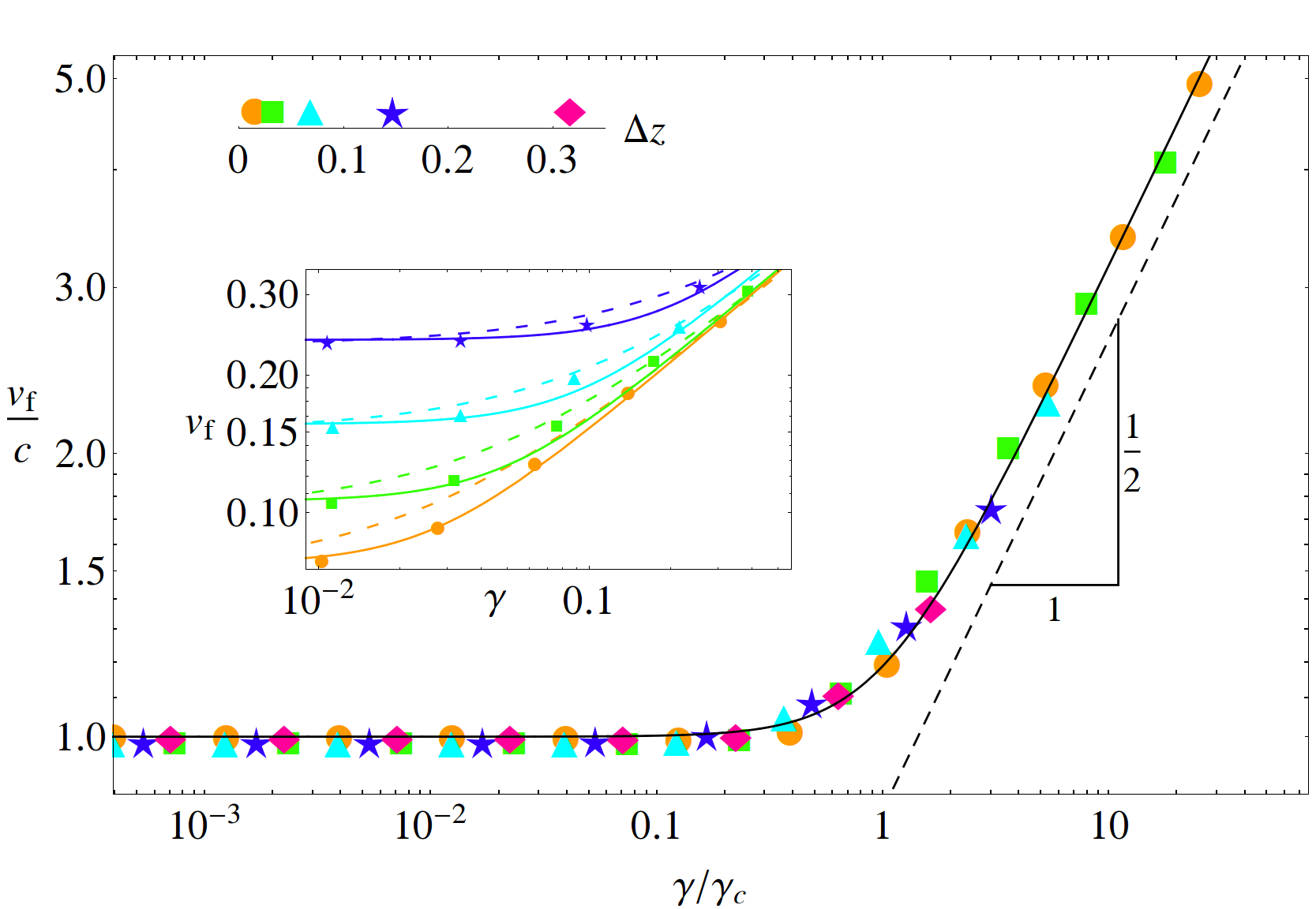}}
    \put(32,180){(a)}
  \end{picture}
   \centering
 \end{minipage}~~
\begin{minipage}[h]{.385\textwidth}
  \begin{picture}(220,100)
    \put(0,0){\includegraphics[width=\textwidth]{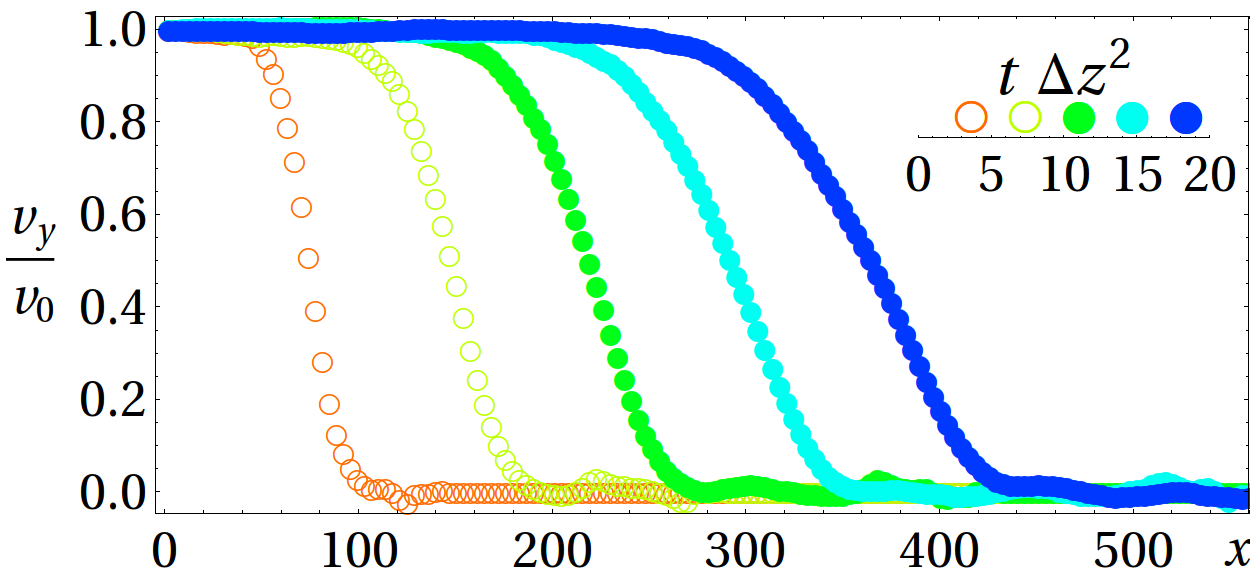}}
    \put(26,75){(b)}
  \end{picture}\\ \vspace{-0pt} 
  \begin{picture}(220,100)
    \put(0,0){\includegraphics[width=1\textwidth,trim=0 0 0 60, clip=true]{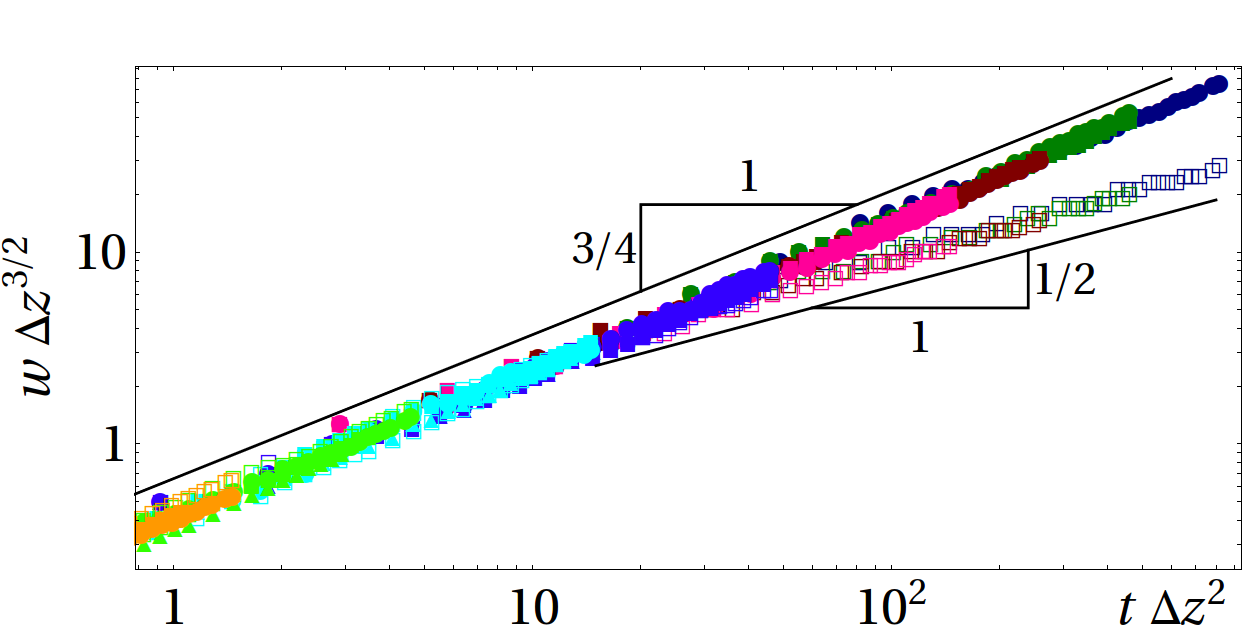}}
    \put(26,80){(c)}
  \end{picture}
   \centering
 \end{minipage} 
\centering
\caption{(a) In the main panel, the normalized front velocity, $v_f/c$, is plotted as a function of the normalized strain $\gamma/\gammac$ for a range of $\dz$. For $\gamma/\gammac \ll 1$, the front speed $v_f$ is independent of the applied strain and corresponds to the linear speed of sound $c$. In the strongly non-linear regime $\gamma/\gammac \gg 1$, the front speed is independent of $\dz$ and scales with the applied strain as $\vf\propto\gamma^{1/2}$ (straight black line). The solid curve is plotted using the relation $\vf = {(\Eo^2 + {k_{nl}}^2 \gamma^2)^{1/4}}/{\sqrt{\rho}}$. The inset compares the solid curves from the main plot to the alternative relation $\vft = \sqrt{c^2 + \frac{k_{nl} \gamma}{\rho}}$ (dashes lines), clearly favoring $\vf$ over $\vft$.
(b) Evolution of a non-linear ($\gamma/\gammac = 6.8$) wave front for $\dz \approx 0.15$. Late times $t \dz^2 > 10$ are indicated by full circles. 
(c) Time dependence of the widths in the non-linear regime. Different colors correspond to different $\dz$, as in the plot on the left. The full symbols correspond to the non-linear regime with $1.5<\gamma/\gammac <680$. The open symbols correspond to the linear regime ($\gamma/\gammac < 10^{-3}$). } 
\label{fig:VelVsGamma}
\end{figure*}
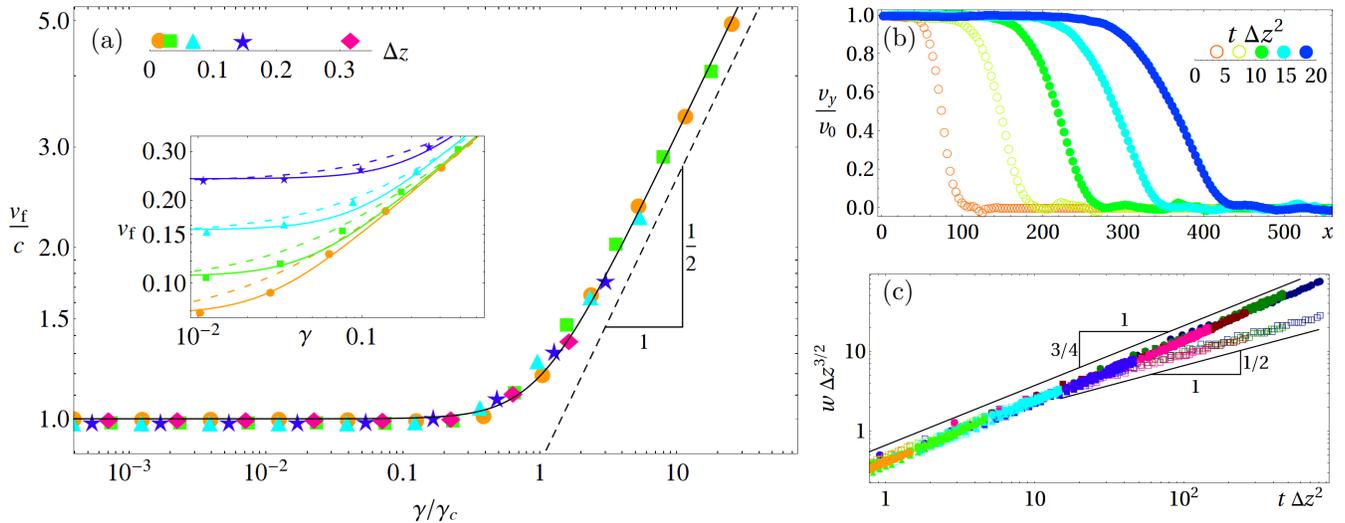

Upon defining a non-linear shear modulus $G_{nl} \equiv \sigma/\gamma=G+k_{nl}\gamma^{n-1}$, we can obtain the characteristic non-linear front speed $\vf$ as
\begin{equation}
\vf = \sqrt{\frac{G_{nl}}{\rho}} =  \sqrt{ c^2 + \frac{ k_{nl} \gamma^{n-1}}{\rho}} \;. \label{nl_speed}
\end{equation}
Notice that for $\gamma\ll\gammac$, $\vf$ approaches the transverse linear speed of sound $c=\sqrt{{G}/{\rho}}\propto \dz^{{\beta}/{2}}$. For $\gamma\gg\gammac$, instead, $\vf \propto \gamma^{\frac{n-1}{2}}$ becomes independent of $\dz$ and controlled only by the applied strain: the hallmark of a strongly non-linear wave.  

In the main panel of Fig.\ 5 (a), we plot $v_\text{f}/c$ versus $\gamma/\gamma_c$ for homogeneously cut networks of various $\dz$ and achieve a very good data collapse.  For $\gamma \gg \gammac$, we find $\vf \propto \gamma^{{1}/{2}}$ (i.e.\ $n=2$), which implies that $\gammac \propto \dz$ (recall $\beta \approx 1$). Hence, the linear sound regime $\gamma \ll \gammac$, where $\vf \approx c$, progressively shrinks to zero as the critical point is approached. Furthermore, the (cutting-protocol independent) finding $n=2$ and mirror symmetry imply that the stress attains the surprising non-analytic form $\gamma |\gamma|$ at the critical point \cite{Wyart2008,old}. We find that the simple analytic expression $\tilde{v}_f = \left(G^2 + {k_{nl}}^2 \gamma ^2\right)^{1/4}$, plotted as a continuous line in the inset of Fig.\ 5 (a), fits our data very well for {\it all} $\dz$ and $\gamma$.   

Since the dynamics of random spring networks is inherently over-damped, we do not observe a stationary shock solution. Instead the shock width broadens as it propagates, as shown in Fig.\ 5 (b). However, there is a striking difference with the linear case which is illustrated in Fig.\ 5 (c). In the highly non-linear regime, the shock width (closed symbols) broadens {\it super-diffusively} as $w \propto t^{3/4}$ even at late-times $t \gg t_c$, when the linear fronts (open symbols) have clearly crossed-over to {\it diffusive} broadening. This means that the quasi-static approximation, from which the late-times diffusive broadening was derived, ceases to be valid in the non-linear regime. 

The resilience of the super-diffuive broadening in the non-linear regime is one of the central results of our work: it can be explained qualitatively by an energy cascade mechanism from low to high $\omega$, reminiscent of acoustic turbulence. When large oscillatory strains with frequency components $\omega \ll \omega_c$ are applied to the network, non-linearities act as an additional source that generates higher and higher harmonics, past the characteristic threshold $\omega_c$. These higher harmonics keep the network away from attaining the quasi-static limit. Consequently, one of the tangible experimental signatures of this energy cascade is the persistence of the regime of super-diffusive broadening even at long times when diffusive broadening is observed for linear fronts, see Fig.\ 5 (c) \footnote{The constant shear we apply excites simultaneously all $\omega$ components.}.          

To sum up, we find that at (or near) the isostatic point of a random network of masses permanently connected by Hookean springs (a) shear fronts propagate as supersonic shocks with speed $\vf \propto |\gamma|^{1/2}$ and (b) their widths broaden super-diffusively as $t^{3/4}$. We propose that the inherent non-linearities at the critical point trigger an {\it energy cascade} from low to high frequency components that causes the breakdown of the quasi-static approximation and gives rise to super-diffusion. In the linear regime, we observe at late times a {\it diffusive} broadening controlled by an effective shear viscosity that diverges at the critical point.  Even in the limit of vanishing microscopic coefficient of dissipation, (strongly) disordered networks behave as if they were {over-damped} because energy is irreversibly leaked into diverging non-affine fluctuations. The shear front rheology approach illustrated here can be used more broadly to infer the viscoelastic properties of any material undergoing very slow dynamics from (non-linear) acoustic experiments.   

\begin{acknowledgments}
We wish to acknowledge helpful conversations with L. Gomez, B. Tighe, A. Zaccone, M. van Hecke, B. Tighe, M. Wyart, P. Hebraud and F. Lequeux.
This work was supported by FOM and NWO.
\end{acknowledgments}

\section{\label{appA} Supplementary Information: from oscillatory to shear front rheology}

In this appendix we demonstrate the equivalence between shear front rheology and oscillatory rheology. Consider the frequency dependent constitutive relation for a linear visco-elastic material
\begin{equation}
\sigma(s) = \E(s) \gamma(s) \;, \label{eq:sigmaGeneralLaplace}
\end{equation}
where,  $G(s)_{s=i\omega}=G'(\omega) + iG''(\omega)$ is the (Laplace transformed) complex modulus, whose real and imaginary parts correspond to the storage and loss modulus respectively. The field  $\sigma(s)\equiv \sigma_{xy}(s)$ denotes the shear stress and $\gamma(s)\equiv\gamma_{xy}(s)$ the shear strain. 
The general frequency dependent constitutive stress-strain relation Eq.\ (\ref{eq:sigmaGeneralLaplace}) corresponds to the following convolution integral in the time domain,
\begin{eqnarray}
\sigma(x,t) =  \int_0^t G(\tau)\gamma(x,t-\tau)d\tau. \label{stress-strain} 
\end{eqnarray}
Substituting in the equation of motion $\frac{\partial\sigma(x,t)}{\partial x}= \rho\frac{\partial^2 u(x,t)}{\partial t^2}$ and defining $v(x,t)=\frac{\partial u(x,t)}{\partial t}$, we obtain (interchanging the order of integration and differentiation),
\begin{eqnarray}
\int_0^t G(\tau)\frac{\partial\gamma(x,t-\tau)}{\partial x} d\tau= \rho\frac{\partial v(x,t)}{\partial t}.
\end{eqnarray}
Differentiating with respect to time $t$, we obtain
\begin{eqnarray}
\left[G(\tau)\frac{\partial\gamma(x,t-\tau)}{\partial x}\right]_{\tau=t}  &+ \int_0^t G(\tau)\frac{\partial^2\gamma(x,t-\tau)}{\partial x\partial t} d\tau &= \nl \rho\frac{\partial^2 v(x,t)}{\partial t^2}.   
\end{eqnarray}
Since $\gamma(x,t)=\frac{\partial u(x,t)}{\partial x}$, the first term in the square brackets can be expressed as $G(t)\frac{\partial^2 u(x,0)}{\partial x^2}$, which evaluates to 0, since there is no displacement field at $t=0$. Therefore, 
\begin{eqnarray}
\frac{\partial^2}{\partial x^2}\int_0^t G(\tau)v(x,t-\tau) d\tau= \rho\frac{\partial^2 v(x,t)}{\partial t^2}.
\end{eqnarray}
 Transforming back to Laplace time (noting that the first term is just the convolution integral), we obtain
\begin{eqnarray}
G(s)\frac{\partial^2 v(x,s)}{\partial x^2} &= \rho s^2v(x,s)-\rho  s\left[v(x,t)\right]_{t=0} &- \nl \rho\left[\frac{\partial v(x,t)}{\partial t}\right]_{t=0}
\end{eqnarray}
Since we are solving in the domain $x>0$, our initial condition is $v(x>0,0)=0$ and boundary condition is $v(x=0,t)=v_0$. Thus, the two terms on the right are 0 leaving us with
\begin{eqnarray}
\frac{\partial^2 v(x,s)}{\partial x^2}= \frac{\rho s^2}{G(s)}v(x,s).
\end{eqnarray}
This  is an ordinary differential equation in $x$, which is solved to obtain Eq.\ (\ref{eq:app:v(x,s)}).
\\
\\
In the quasi-static regime, $G(s)= G+s\eta$ \cite{tighe2011}. Non-dimensionalizing Eq.\  (\ref{eq:app:v(x,s)}) with the characteristic time scale $t^*=\frac{\eta}{G}$ and length scale $x^*=\frac{\eta}{\sqrt{ G \rho }}$, we obtain
\begin{align}
 \frac{\V(\X,\S)}{\Vo} = \frac{1}{\S} \exp\!\left( -\frac{ \S \X }{\sqrt{1+\S}}  \right) \; , \label{eq:app:V(X,S)}
\end{align}
expressed in terms of hatted variables, eg. $\X=\frac{x}{x^*}$ and $\S=s t^*$. After expanding in a Taylor series, one obtains, 
\begin{align}
 \frac{\V(\X,\S)}{\Vo}  = \frac{1}{\S} \sum_{n=0}^{\infty} \frac{1}{n!} \left( - \frac{ \X \S }{ \sqrt{1+\S} } \right)^{\!n} \;.
\end{align}
Taking the inverse Laplace transform term by term, 
\begin{align}
 \frac{\V(\V,\T)}{\Vo}  = \sum_{n=0}^{\infty} \frac{\X^n}{n!} \mathcal{L}^{-1} 
    \left[ \frac{ (-1)^n \S^{n-1} }{ (1+\S)^{n/2} } \right] \;.
\end{align}
we obtain the complete solution
\begin{equation}
 \frac{\V(\X,\T)}{\Vo} = \sum_{n=0}^{\infty} \frac{(\X \T^{-1/2})^n}{n!}~_1\!\tilde F_1(n/2,1\!-\!n/2;-\T) \; \label{eq:apndx:V(XT)1F1},
\end{equation}
where  $_1\!\tilde F_1(a,b;z)$ is the regularized hypergeometric function \cite{1F1regularized}. Equation (\ref{eq:apndx:V(XT)1F1}) is used to plot the continuous line that fits the numerical data for the velocity profiles in the main panel of Fig.\ 2 .

In the long time limit, we obtain the following  asymptotic form 
\begin{eqnarray}
 \frac{\V(\X,\T)}{\Vo} = \frac{1}{2} \erfc\!\left( \frac{\X - \frac{1}{2} - \T}{2\sqrt{\T}} \right)  &\text{for } \T \to \infty .
\end{eqnarray}
\\
\\
At higher frequencies (or equivalently shorter times), we assume that the complex modulus crosses over to a power law regime that for the homogeneously cut networks reads $G(s)\propto s^{1/2}$ \cite{tighe2011}. Since we do not need the explicit analytical solution in this regime, we obtain the functional form of the solution from the following general properties of the Laplace transform. For a given function $f(t)$, let $\hat f(s) := \mathcal{L}[f(t)](s)$ be its Laplace transform. Then
\begin{equation}
 \mathcal{L}[f(x^{-\beta} t)](s) = x^\beta \hat f(x^\beta s) \;,
\end{equation}
which can  be shown by a change of variables in the definition of the Laplace transformation. Assume $\E(s) \propto s^\alpha$ and note that the velocity profile, Eq.~\eqref{eq:app:v(x,s)}, can be written as:
\begin{align}
 \frac{v(x,s)}{v_0} = \frac{1}{s} \exp( - C s^{1-\alpha/2} x) \;,
\end{align}
where the constant $C$ accounts for the proportionality factor in $\E(s)$. Hence, with $1/\beta = 1-\alpha/2$:
\begin{align}
 \frac{v(x,s)}{v_0} = x^\beta \frac{\exp\!\big( - C (x^\beta s )^{1/\beta} \big)}{x^\beta s} =  x^\beta \hat f(x^\beta s)  \;,
\end{align}
where $\hat f(s) = {\exp( - C s^{1/\beta} )}/s$. Consequently, in real time, the velocity profile must obey the scaling relation $v(x,t) = f(x^{-\beta} t)$, which can be rewritten as:
\begin{equation}
 v(x,t) = F\left( \frac{x}{w(t)} \right) \quad \text{where} \quad w(t) \propto t^{1-\alpha/2} \;.
\end{equation}
For our case, $\alpha = 1/2$, we thus expect the front to emerge from the boundary with a width $w_d(t) \propto t^{3/4}$ as we found in the main text. 

\bibliographystyle{unsrt}
\bibliography{SoftMatter}
\end{document}